\documentclass{elsart}
\usepackage{natbib}
\usepackage{epsfig}
\begin{document}
\runauthor{Tacconi-Garman}
\begin{frontmatter}
\title{SINFONI Observations of Starclusters in Starburst Galaxies}
\author[ESO]{Lowell E. Tacconi-Garman\thanksref{X}}
\thanks[X]{On behalf of the SINFONI Science Verification Starburst Team}

\address[ESO]{ESO, Karl-Schwarzschild-Stra\ss e 2, D-85748 Garching
bei M\"unchen, Germany}
\begin{abstract}
We have used ESO's new NIR IFS SINFONI during its Science
Verification period to observe the central regions of local starburst
galaxies. Being Science Verification observations, the aim was
two-fold: to demonstrate SINFONI's capabilities while obtaining
information on the nature of starclusters in starburst galaxies. The
targets chosen include a number of the brighter clusters in NGC1808
and NGC253. Here we present first results.
\end{abstract}
\end{frontmatter}

\typeout{SET RUN AUTHOR to \@runauthor}

\section{Introduction}
That star formation in starburst galaxies manifests itself in the
production of numerous young, massive, clusters or super-starclusters
is by now observationally quite well established (e.g.~\cite{TGSE};
\cite{Mengel05}
and references therein; \cite{deGrijs}).  At least in part owing
to the strong, patchy obscuration in these systems the nature of these
clusters is less well known, especially with respect to their age, mass,
boundedness and the likelihood of their survival (but see, for example,
\cite{Mengel05} and \cite{Gieles05}).  We have made short observations
of near- and
circumi-nuclear fields in two nearby starburst galaxies, NGC\,253 and
NGC\,1808, during SINFONI Science Verication.  We are using these data
to better characterize the central starburst regions and starclusters
in these systems.  Here we present a brief summary of initial results.

\section{SINFONI and Observation Overview}

SINFONI, offered to the community since 01~Apr.~2005, is a
combination of a NIR Integral Field Spectrometer (SPIFFI)
and a MACAO adaptive
optics module.  Briefly, SPIFFI, uses
gratings to provide the user with data with a spectral
resolution of 2000, 3000, 4000 (in the J, H, and K~Bands,
respectively), or 1500 for the combined H+K Bands.  The spatial
sampling of 64$\times$32 pixels over square fields-of-view
0.8$^{\prime\prime}$, 3$^{\prime\prime}$, or 8$^{\prime\prime}$ on a
side is achieved by slicing the field with a mirror-pair image slicer
and imaging it onto a Hawaii 2RG detector.  The largest of these
fields-of-view is meant primarily seeing-limited operation, as is the
case in the absence of a suitable adaptive optics reference target.

The MACAO unit is very broadband in its response, though is peaked
in the R~Band.  Although targets with R~magnitudes less than about
11 yield the best corrections, some correction can be achieved to
R$\sim$17 under the best conditions.  In very good conditions moderate
improvement can be achieved even to 30$^{\prime\prime}$ separation
between the AO-target and the science target\footnote{Further
details for SINFONI can be found on the ESO SINFONI webpage, {\tt
http://www.eso.org/instruments/sinfoni/index.html}, and on pages linked
to from that page.}.

The observations reported on here were made with the H+K
grating and the 0.8$^{\prime\prime}$ (NGC\,253; two fields with 200
and 480~sec exposures) and
0.8$^{\prime\prime}$ (NGC\,1808; two fields with 10 and 20~min
exposures) pixel scales.  The NGC\,1808 observations used the MACAO
unit, while the NGC\,253 observations were done in seeing-limited
mode.

\section{First Results}

\subsection{Br$\gamma$ and He{\small I} in NGC\,253}

In Figure~1 we present images of the log of the Br$\gamma$ emission
(left),
the log of the equivalent width of Br$\gamma$ (center), and the ratio of the 
2.058$\,\mu$m 2$^1$P--2$^1$S He{\small I} line to
Br$\gamma$ (right) in the circumnuclear region of NGC\,253.  The
Br$\gamma$ is seen to peak strongly at the position of the well-known
IR peak in this galaxy to the SW on the nucleus, consistent with what
is seen in Pa$\alpha$, Br$\delta$, and 2.058$\,\mu$m 2$^1$P--2$^1$S
He{\small I} (not shown).

\begin{figure}[h!]
\begin{center}
\psfig{figure=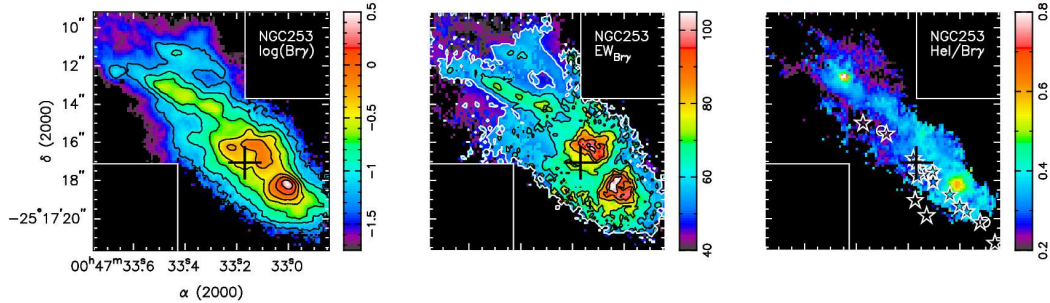, width=13.92cm}
\end{center}
\caption{\label{NGC253.Brg.fig}Images of the circumnuclear region of
NGC\,253 as described in the text.  In each panel the black {\bf +}
indicates the position of the nucleus, and the white lines indicate
the limits of the observed mosaic.  In the right panel circles
indicate ultracompact H{\small II} regions from \cite{Johnson2001}, stars
are 2\,cm radio continuum sources from \cite{UA97};
see \cite{TGSE} for details.}
\end{figure}

Although the 2.058$\,\mu$m 2$^1$P--2$^1$S He{\small I} line
is not an ideal tracer of the hardness of the radiation field
(\cite{Lumsden2001},\cite{RigbyRieke04}), its ratio with Br$\gamma$ and
the EW$_{\mathrm Br\gamma}$ in the two peaks shown in Figure~1 (right)
are consistent with starburst age of ~5.5\,Myr with the most massive stars
being $\sim$30\,M$_\odot$.  This is less massive than postulated by some
authors for the bright peak to the SW of the nucleus, but the resulting
radiation field is likely consistent with what would be required to
explain the strong decrease in the 3.3$\,\mu$m PAH feature-to-continuum
ratio at these positions \cite{PAH}.  Observations of a ``pure''
line\footnote{One for which the line strength is determined almost
solely from the recombination cascade physics.}
of He{\small I}, such as the 1.7007\,$\mu$m 4$^3$D--3$^3$P
line \cite{Lumsden2001}, would
be useful to properly constrain this scenario.  

\subsection{A Comparison with {\em HST\/} NICMOS}

As a means of demonstrating the combined capabilities afforded by both
the VLT-UT4 and SINFONI itself, we present in Figure~2 an image of the
1.644$\,\mu$m [Fe{\small II}] line emission from NGC\,253 as observed
with SINFONI\@.  This region has been previously observed in the same
line with {\em HST\/} NICMOS \cite{Alonso-Herrero2003}, and the
contours in Figure~2 are derived from that data.  The correspondence
is quite remarkable indeed, especially considering that the SINFONI
data are seeing-limited.  However, the true advantage of SINFONI over
{\em HST\/} NICMOS lies in the wealth of information afforded by the
IFS itself (through velocity (dispersion) maps, not shown here due to
space limitations).

\begin{figure}[h!]
\begin{center}
\psfig{figure=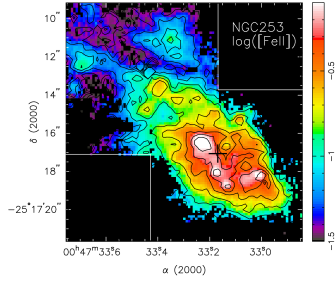, width=4.38cm}
\end{center}
\caption{\label{NGC253.FeII.fig}An image of the 1.644$\,\mu$m
[Fe{\small II}] emission in the circumnuclear region of NGC\,253
observed with SINFONI on the VLT-UT4\@.  Markings are as in Figure~1.
{\em The overlaid contours are derived from the {\em HST\/} 
NICMOS observations of \cite{Alonso-Herrero2003}.}}
\end{figure}

\subsection{Star clusters in NGC\,1808}

The data from our AO-assisted observations of two clusters in
NGC\,1808 are a bit noisier than those of NGC\,253, which compromises
(at least thus far) our ability to construct
fruitful line maps.  However, in Figure~3 we present collapsed K-Band
images of the fields observed, with each of the two right panels being
0.8$^{\prime\prime}$ on a side.  We have used our best guess at the
PSF (from the telluric calibration star) and infer an upper limit on
the size of these clusters of 15\,pc and 8\,pc for Knots~20 and 12
(Figure~3 top right, and bottom right), respectively.  This is broadly
consistent with the sizes quoted in \cite{TGSE}.
\begin{figure}
\begin{center}
\psfig{figure=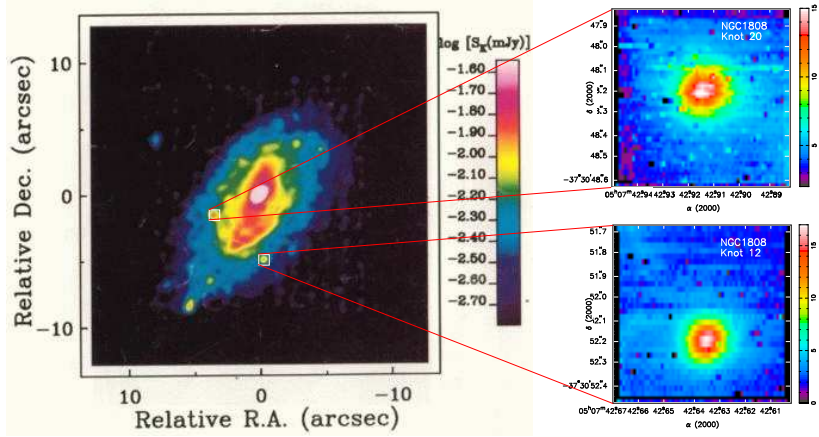, width=10.80cm}
\end{center}
\caption{\label{NGC1808_proceedings.xfig}(left) K-Band speckle image of the
circumnuclear region of NGC\,1808 \cite{TGSE} illustrating the SSC knots, 
(right) preliminary reductions of 
SINFONI observations of two of the clusters represented as data cubes
collapsed into K-Band images.}
\end{figure}

\end{document}